\documentclass[aps,onecolumn,letterpaper,oneside,preprint,tightenlines,draft,%
nobibnotes,nofootinbib,amsfonts,amssymb,amsmath,showpacs,%
showkeys]{revtex4}

\newcommand{\cN}{\mathcal{N}}
\newcommand{\cP}{\mathcal{P}}
\newcommand{\cT}{\mathcal{T}}
\newcommand{\cV}{\mathcal{V}}
\newcommand{\cW}{\mathcal{W}}
\newcommand{\bH}{\boldsymbol{H}}
\newcommand{\bQ}{\boldsymbol{Q}}
\newcommand{\tH}{\tilde{H}}

\newcommand{\tcV}{\tilde{\cV}}
\newcommand{\bbC}{\mathbb{C}}
\newcommand{\hA}{\widehat{A}}
\newcommand{\hP}{\widehat{P}}
\newcommand{\hQ}{\widehat{Q}}
\newcommand{\hE}{\widehat{E}}
\newcommand{\hW}{\widehat{W}}
\newcommand{\rmi}{\mathrm{i}}
\newcommand{\rme}{\mathrm{e}}
\newcommand{\rmd}{\mathrm{d}}

\newcommand{\rmA}{\mathrm{A}}
\newcommand{\rmP}{\mathrm{P}}
\newcommand{\csch}{\operatorname{csch}}

\begin{document}



%
%

\title{Existence of Different Intermediate Hamiltonians in Type A
 $\cN$-fold Supersymmetry}
\author{Bijan Bagchi}
\email{bbagchi123@rediffmail.com}
\affiliation{Department of Applied Mathematics, University of Calcutta,\\
 92 Acharya Prafulla Chandra Road, Kolkata 700 009, India.}
\author{Toshiaki Tanaka}
\email{ttanaka@mail.ncku.edu.tw}
\affiliation{Department of Physics, National Cheng Kung University,\\
 Tainan 701, Taiwan, R.O.C.\\
 National Center for Theoretical Sciences, Taiwan, R.O.C.}


\begin{abstract}

Type A $\cN$-fold supercharge admits a one-parameter family of
factorizations into product of $\cN$ first-order linear differential
operators due to an underlying $GL(2,\bbC)$ symmetry. As a consequence,
a type A $\cN$-fold supersymmetric system can have different
intermediate Hamiltonians corresponding to different factorizations.
We derive the necessary and sufficient conditions for the latter
system to possess intermediate Hamiltonians for the $\cN=2$ case.
We then show that whenever it has (at least) one intermediate
Hamiltonian, it can admit second-order parasupersymmetry and a
generalized 2-fold superalgebra. As an illustration, we construct
a set of generalized P\"{o}schl--Teller potentials of this kind.

\end{abstract}






\maketitle

\section{Introduction}
\label{sec:intro}

Supersymmetric quantum mechanics (SUSY QM) has constituted an active
research field in theoretical sciences over two decades. Although
the original motivation for studying SUSY QM was to unveil the
mechanisms of its dynamical breaking in quantum field
theories~\cite{Wi81}, it turned out that SUSY QM, as the minimum
building block of SUSY, contains various relevant concepts
which provide convenient platforms to uncover many useful properties
of quantum mechanics~\cite{CKS95,Ju96,Ba00}. In particular, it is
consistent with factorization schemes~\cite{Sc40a} and intertwining
relationships~\cite{Da1882} thereby providing a powerful tool to
construct solvable Schr\"{o}dinger equations. Furthermore, interesting
extensions to higher-order SUSY schemes were carried out by taking
recourse to higher-derivative versions of the factorization
operators~\cite{AIS93,AST01b,AS03}.

The original extension in Ref.~\cite{AIS93} was considered by using
higher-order intertwining operators which are expressed as products
of first-order linear differential operators and then applying
the ordinary SUSY results. Later in Refs.~\cite{AICD95,AIN95a},
through the analysis of the general second-order case, the concept
of \emph{reducibility} was introduced. In this regard,
a higher-order intertwining operator is said to be \emph{reducible}
if it is factorized into a product of first-order differential
operators such that with respect to each factor there exists
an intermediate \emph{real} Hamiltonian satisfying a (shifted) SUSY
relation. Otherwise it is called \emph{irreducible}. This concept,
however, seems less useful in view of the current status where
non-Hermitian quantum theories have been investigated intensively
since the discovery of $\cP\cT$ symmetry~\cite{BB98a}. In fact,
the SUSY method turned to be useful also in constructing
a \emph{complex} potential with real
spectrum~\cite{CJT98,BR00,ZCBR00}.

In addition to the usefulness of the reality constraint, there
arises a natural question about the well-definiteness of the concept
if we take into account the fact that in general factorization of
higher-order linear differential operators is not unique. The latter
fact indeed has been reported in the context of the factorization
method and ordinary SUSY QM, that is, there exist several
Schr\"{o}dinger operators that admit different
factorizations~\cite{Mi84,Fe84,Zh87,Ku87,AF88,Fi88,MRLB89}, see
also Ref.~\cite{Qu08b} for a recent approach.
Hence, a reducible higher-order intertwining operator may admit
simultaneously another factorization for which there are no
intermediate real Hamiltonians.

On the other hand, the non-uniqueness in factorizing intertwining
operators of arbitrary finite orders was, to the best of our
knowledge, first reported in Ref.~\cite{AST01a} for the well-known
quasi-solvable sextic anharmonic oscillator potentials in the framework
of type A $\cN$-fold supersymmetry (see, Eq.~(47) in the latter
reference). Later it was shown that the non-uniqueness of
factorizations in type A $\cN$-fold SUSY is a consequence of
the underlying $GL(2,\bbC)$ symmetry~\cite{Ta03a}.

In a recent communication~\cite{BQR08}, the following two-parameter
family of second-order supersymmetric (SSUSY) system $(h^{(1)},h^{(2)},
\hat{A}\hat{B})$, characterized by the two parameters $A$ and $B$,
satisfying $\hat{A}\hat{B}h^{(1)}=h^{(2)}\hat{A}\hat{B}$ was constructed:
\begin{align}
h^{(1)}&=\hat{B}^{\dagger}\hat{B}+\frac{\bar{c}}{2}
 =-\frac{\rmd^{2}}{\rmd x^{2}}+V_{A,B}(x)-\tilde{E}+\frac{\bar{c}}{2},
\label{eq:BQRs}\\
h^{(2)}&=\hat{A}\hat{A}^{\dagger}-\frac{\bar{c}}{2}
 =-\frac{\rmd^{2}}{\rmd x^{2}}+V_{A,B,\text{ext}}(x)-E
 -\frac{\bar{c}}{2},
\end{align}
where
\begin{align}
V_{A,B}(x)&=[B^{2}+A(A+1)]\csch^{2}x-B(2A+1)\csch x\coth x,\\
V_{A,B,\textrm{ext}}(x)&=V_{A,B}(x)+\frac{2(2A+1)}{2B\cosh x-2A-1}
 -\frac{2[4B^{2}-(2A+1)^{2}]}{(2B\cosh x-2A-1)^{2}},
\end{align}
and the constants $E$, $\tilde{E}$, and $\bar{c}$ are given by
\begin{align}
E=-\left(B\mp\frac{1}{2}\right)^{2},\quad
 \tilde{E}=-\left(B\pm\frac{1}{2}\right)^{2},\quad\bar{c}=\mp2B.
\end{align}
The intertwining operators $\hat{A}$ and $\hat{B}$ are respectively
given by
\begin{align}
\hat{A}&=\frac{\rmd}{\rmd x}\pm\left(B\pm\frac{1}{2}\right)\coth x
 \mp\left(A+\frac{1}{2}\right)\csch x-\frac{2B\sinh x}{2B\cosh x-2A-1},\\
\hat{B}&=\frac{\rmd}{\rmd x}\mp\left(B\pm\frac{1}{2}\right)\coth x
 \pm\left(A+\frac{1}{2}\right)\csch x.
\end{align}
It was further shown in Ref.~\cite{BQR08} that the system admits the
intermediate Hamiltonians $h$ given by
\begin{align}
h&=\hat{A}^{\dagger}\hat{A}-\frac{\bar{c}}{2}
 =-\frac{\rmd^{2}}{\rmd x^{2}}+V_{A,B\pm1}(x)
 -E-\frac{\bar{c}}{2}\notag\\
&=\hat{B}\hat{B}^{\dagger}+\frac{\bar{c}}{2}
 =-\frac{\rmd^{2}}{\rmd x^{2}}+V_{A,B\pm1}(x)
 -\tilde{E}+\frac{\bar{c}}{2},
\end{align}
with
\begin{align}
V_{A,B\pm1}(x)=[(B\pm1)^{2}+A(A+1)]\csch^{2}x
 -(B\pm1)(2A+1)\csch x\coth x,
\label{eq:BQRe}
\end{align}
which satisfy the ordinary SUSY relations $\hat{A}h=h^{(2)}\hat{A}$ and
$\hat{B}h^{(1)}=h\hat{B}$. We shall hereafter call the system
(\ref{eq:BQRs})--(\ref{eq:BQRe}) the \emph{BQR SSUSY model}.

Taking into account the fact that type A 2-fold SUSY is the necessary
and sufficient condition for the existence of (at least) two linearly
independent analytic (local) solutions to Schr\"{o}dinger equation of
one degree of freedom~\cite{GT06}, we immediately know that the above
BQR SSUSY model, which is exactly solvable, also belongs to type A
2-fold SUSY. Therefore, it would be natural that it has (at least)
two different factorizations of the second-order intertwining operator
in view of the aforementioned $GL(2,\bbC)$ symmetry and has two
different intermediate Hamiltonians correspondingly.

Regarding the existence of intermediate Hamiltonians, we note
the following fact shown in Ref.~\cite{AST01a} that
the most general form of type A $\cN$-fold SUSY quantum systems
constructed directly from the $\cN$th-order intertwining operators
of type A, namely, type A $\cN$-fold supercharge
$P_{\cN}^{-}=P_{\cN 1}^{-}\dots P_{\cN\cN}^{-}$ (cf.,
Eq.~(\ref{eq:ANfch})) by solving $P_{\cN}^{-}H^{-}=H^{+}P_{\cN}^{-}$
is more general than that of the systems constructed from the $\cN$
repeated applications of the first-order intertwining operators
$P_{\cN k}^{-}$ by solving $P_{\cN k}^{-}H^{(k-1)}=H^{(k)}P_{\cN k}^{-}$
($k=1,\dots,\cN$) with the identification $H^{-}=H^{(0)}$ and
$H^{+}=H^{(\cN)}$. It is apparent that in the latter construction we
automatically obtain a series of the intermediate Hamiltonians
$H^{(1)},\dots,H^{(\cN-1)}$ in addition to the $\cN$-fold SUSY pair
Hamiltonians $H^{\pm}$ at the cost of the generality. In the former
construction, on the other hand, the existence of intermediate
Hamiltonians is not guaranteed in general. Hence, the fact that the
type A $\cN$-fold supercharge has the factorized form by definition
does not necessarily imply the existence of intermediate Hamiltonians. 
In particular, the fact that type A $\cN$-fold supercharge admits
different factorizations does not automatically mean that the most
general type A $\cN$-fold SUSY system has different sets of
intermediate Hamiltonians accordingly.

Motivated by the backgrounds described above, we investigate in this
article under what conditions type A $\cN$-fold SUSY systems admit
intermediate Hamiltonians in the case of $\cN=2$. Furthermore, we
also examine under the satisfaction of the conditions how many sets
of such Hamiltonians are admissible for the type A 2-fold SUSY
systems. In addition, we show that any such a system has another
type of nonlinear supersymmetries, namely, parasupersymmetry
of order 2~\cite{RS88}.

We organize the article as follows. In the next section, we review
the framework of type A $\cN$-fold SUSY by putting emphasis on the
$GL(2,\bbC)$ symmetry. In Section~\ref{sec:inHam}, we investigate
in details under what conditions a type A 2-fold SUSY quantum system
has one or more intermediate Hamiltonians. In particular, we show that
the maximum number of different intermediate Hamiltonians in type A
2-fold SUSY is two. In Section~\ref{sec:psusy}, we further show that
when a type A 2-fold SUSY system admits (at least) one intermediate
Hamiltonian, the system can have second-order parasupersymmetry.
A novel generalization of 2-fold superalgebra is discussed briefly.
As an application of the results, we construct in
Section~\ref{sec:appli} a type A 2-fold SUSY system with two
different intermediate Hamiltonians of generalized
P\"{o}schl--Teller type which includes the BQR SSUSY model as
a particular case. Then, we close the article with discussion and
perspectives of further developments in the last section.

\section{Type A $\cN$-fold Supersymmetry and $GL(2,\bbC)$ Covariance}
\label{sec:typeA}

Roughly speaking, type A $\cN$-fold SUSY quantum systems are composed
of a pair of scalar Hamiltonians $H^{\pm}$ and an $\cN$th-order
linear differential operator $P_{\cN}^{-}$ of the following
forms:\footnote{We keep the original notations as far as possible.
Thus, note that the function $E(x)$ is different from the constant
$E$ in the BQR SSUSY model (\ref{eq:BQRs})--(\ref{eq:BQRe}).
Similarly, the function $A(z)$ introduced later in (\ref{eq:gHams})
is different from the parameter $A$ in the latter model.}
\begin{gather}
H^{\pm}=-\frac{1}{2}\,\frac{\rmd^{2}}{\rmd x^{2}}+\frac{1}{2}W(x)^{2}
 -\frac{\cN^{\,2}-1}{24}\left(2E'(x)-E(x)^{2}\right)\pm\frac{\cN}{2}
 W'(x)-R,
\label{eq:AHams}\\
P_{\cN}^{-}=\prod_{k=0}^{\cN-1}\left(\frac{\rmd}{\rmd x}+W(x)+\frac{
 \cN-1-2k}{2}E(x)\right),
\label{eq:ANfch}
\end{gather}
where $R$ is a constant while $E(x)$ and $W(x)$ are analytic functions
satisfying
\begin{align}
\left(\frac{\rmd}{\rmd x}-E(x)\right)\frac{\rmd}{\rmd x}\left(
 \frac{\rmd}{\rmd x}+E(x)\right)W(x)=0\quad\text{for}\quad\cN\geq2,
\label{eq:condW}\\
\left(\frac{\rmd}{\rmd x}-2E(x)\right)\left(\frac{\rmd}{\rmd x}-E(x)
 \right)\frac{\rmd}{\rmd x}\left(\frac{\rmd}{\rmd x}+E(x)\right)
 E(x)=0\quad\text{for}\quad\cN\geq3.
\label{eq:condE}
\end{align}
The product of operators appeared in (\ref{eq:ANfch}) is defined by
\begin{align}
\prod_{k=0}^{n}A_{k}\equiv A_{n}\dots A_{1}A_{0}.
\end{align}
The operators $H^{\pm}$ and $P_{\cN}^{-}$ satisfy an intertwining
relation
\begin{align}
P_{\cN}^{-}H^{-}=H^{+}P_{\cN}^{-}.
\label{eq:inter}
\end{align}
One of the most important features of type A $\cN$-fold SUSY quantum
systems is that the gauged Hamiltonians $\tH^{-}$ and $\bar{H}^{+}$
introduced by
\begin{align}
\bar{\tH}^{\pm}=\rme^{\cW_{\cN}^{\pm}}H^{\pm}\rme^{-\cW_{\cN}^{\pm}},
\qquad\cW_{\cN}^{\pm}(x)=\frac{\cN-1}{2}\int\rmd x\,E(x)\mp\int\rmd x\,
 W(x),
\end{align}
preserve the so-called type A monomial space $\tcV_{\cN}^{(\rmA)}$:
\begin{align}
\bar{\tH}^{\pm}\tcV_{\cN}^{(\rmA)}\subset\tcV_{\cN}^{(\rmA)},\qquad
 \tcV_{\cN}^{(\rmA)}=\bigl\langle 1,z(x),\dots,z(x)^{\cN-1}\bigr\rangle,
\label{eq:Amono}
\end{align}
where the new variable $z(x)$ satisfies
\begin{align}
z''(x)=E(x)z'(x).
\label{eq:defE}
\end{align}
Explicitly, they are given by
\begin{align}
\bar{\tH}^{\pm}=&-A(z)\frac{\rmd^{2}}{\rmd z^{2}}+\left[\frac{\cN-2}{2}
 A'(z)\pm Q(z)\right]\frac{\rmd}{\rmd z}\notag\\
&-\left[\frac{(\cN-1)(\cN-2)}{12}A''(z)\pm\frac{\cN-1}{2}Q'(z)+R\right],
\label{eq:gHams}
\end{align}
where the new functions $A(z)$ and $Q(z)$ are defined by
\begin{align}
2A(z)=z'(x)^{2},\qquad Q(z)=-z'(x)W(x).
\label{eq:defAQ}
\end{align}
The conditions (\ref{eq:condW}) and (\ref{eq:condE}) for type A $\cN$-fold
SUSY are reduced to the following simple forms in terms of $z$:
\begin{align}
\frac{\rmd^{3}}{\rmd z^{3}}Q(z)=0\quad\text{for}\quad\cN\geq2,
\label{eq:condQ}\\
\frac{\rmd^{5}}{\rmd z^{5}}A(z)=0\quad\text{for}\quad\cN\geq3.
\label{eq:condA}
\end{align}
In particular, the condition (\ref{eq:condQ}) indicates that $Q(z)$ is
a polynomial of at most second-degree in $z$ for all $\cN\geq2$:
\begin{align}
Q(z)=b_{2}z^{2}+b_{1}z+b_{0}.
\label{eq:Qofz}
\end{align}
In terms of $z(x)$, the potential terms $V^{\pm}(x)$ of type A $\cN$-fold
SUSY Hamiltonians in (\ref{eq:AHams}) are expressed as
\begin{multline}
V^{\pm}(x)=-\frac{1}{12A(z)}\biggl[(\cN^{\,2}-1)\left(A(z)A''(z)
 -\frac{3}{4}A'(z)^{2}\right)-3Q(z)^{2}\\
\mp 3\cN\bigl(A'(z)Q(z)-2A(z)Q'(z)\bigr)\biggr]-R\biggr|_{z=z(x)}.
\label{eq:VAQ}
\end{multline}
The type A $\cN$-fold SUSY systems (\ref{eq:AHams}) and (\ref{eq:ANfch})
have an underlying symmetry which, as we shall show, plays a central
role in investigating the existence of intermediate Hamiltonians. It is
$GL(2,\bbC)$ linear fractional transformations on the variable $z$
introduced as
\begin{align}
z=\frac{\alpha w+\beta}{\gamma w+\delta}\quad(\alpha,\beta,
 \gamma,\delta\in\bbC,\ \Delta\equiv\alpha\delta-\beta\gamma\neq0).
\label{eq:GL2}
\end{align}
Then, the type A monomial space is invariant under the $GL(2,\bbC)$
transformations induced by (\ref{eq:GL2}):
\begin{align}
\tcV_{\cN}^{(\rmA)}[z]\mapsto\widehat{\tcV}{}_{\cN}^{(\rmA)}[w]=
 (\gamma w+\delta)^{\cN-1}\tcV_{\cN}^{(\rmA)}[z]\Bigr|_{z=\frac{
 \alpha w+\beta}{\gamma w+\delta}}=\tcV_{\cN}^{(\rmA)}[w].
\end{align}
The gauged Hamiltonians $\tH^{-}$ and $\bar{H}^{+}$ are both such linear
differential operators that preserve the type A monomial space, as was
shown in (\ref{eq:Amono}). As a consequence, they are covariant under
the following $GL(2,\bbC)$ transformations:
\begin{align}
\bar{\tH}^{\pm}[z]\mapsto\widehat{\bar{\tH}}{}^{\pm}[w]=(\gamma w
 +\delta)^{\cN-1}\bar{\tH}^{\pm}[z](\gamma w+\delta)^{-(\cN-1)}
 \Bigr|_{z=\frac{\alpha w+\beta}{\gamma w+\delta}},
\end{align}
that is, the transformed operators $\widehat{\tH}{}^{-}$ and
$\widehat{\bar{H}}{}^{+}$ both have the same forms as given in
(\ref{eq:gHams}) with $z$ replaced by $w$ and with $A(z)$ and $Q(z)$
replaced by the transformed functions $\hA(w)$ and $\hQ(w)$ given by
\begin{align}
A(z)\mapsto\hA(w)=\Delta^{-2}(\gamma w+\delta)^{4}A(z)\Bigr|_{z=\frac{
 \alpha w+\beta}{\gamma w+\delta}},\\
Q(z)\mapsto\hQ(w)=\Delta^{-1}(\gamma w+\delta)^{2}Q(z)\Bigr|_{z=\frac{
 \alpha w+\beta}{\gamma w+\delta}}.
\label{eq:traQ}
\end{align}
In particular, the explicit form of $\hQ(w)$ for arbitrary
$\cN\geq2$ is given by
\begin{align}
\hQ(w)=\hat{b}_{2}w^{2}+\hat{b}_{1}w+\hat{b}_{0},
\end{align}
with
\begin{align}
\left(\begin{array}{c}
      \hat{b}_{2}\\ \hat{b}_{1}\\ \hat{b}_{0}
      \end{array}\right)=\Delta^{-1}\left(
 \begin{array}{rrr}
 \alpha^{2} & \alpha\gamma & \gamma^{2}\\
 2\alpha\beta & \alpha\delta+\beta\gamma & 2\gamma\delta\\
 \beta^{2} & \beta\delta & \delta^{2}
 \end{array}\right)\left(
 \begin{array}{c}
 b_{2}\\ b_{1}\\ b_{0}
 \end{array}\right).
\label{eq:trabi}
\end{align}
Utilizing the transformation (\ref{eq:traQ}) and the formulas
\begin{align}
w(x)=-\frac{\delta z(x)-\beta}{\gamma z(x)-\alpha},\qquad
 w'(x)=\frac{\Delta z'(x)}{(\gamma z(x)-\alpha)^{2}}=\Delta^{-1}
 (\gamma w(x)+\delta)^{2}z'(x),
\end{align}
we obtain the transformations of $W(x)$ and $E(x)$ as
\begin{align}
W(x)&\mapsto\widehat{W}(x)=-\frac{\widehat{Q}(w)}{w'(x)}
 =-\frac{Q(z)}{z'(x)}=W(x),
\label{eq:traW}\\
E(x)&\mapsto\widehat{E}(x)=\frac{w''(x)}{w'(x)}=\frac{z''(x)}{z'(x)}
 -\frac{2\gamma z'(x)}{\gamma z(x)-\alpha}
 =E(x)-\frac{2\gamma z'(x)}{\gamma z(x)-\alpha}.
\label{eq:traE}
\end{align}
The invariance of the pair of type A $\cN$-fold SUSY Hamiltonians
$H^{\pm}$ under the $GL(2,\bbC)$ transformations also follows from
a direct application of (\ref{eq:traW}) and (\ref{eq:traE}):
\begin{align}
H^{\pm}[W,E]=H^{\pm}[\hW,\hE],
\end{align}
since $\hW(x)=W(x)$ and from (\ref{eq:defE}) and (\ref{eq:traE}) we have
\begin{align}
2\hE'(x)-\hE(x)^{2}&=2E'(x)-E(x)^{2}.
\label{eq:traE2}
\end{align}
On the other hand, the invariance of the type A $\cN$-fold supercharge
$P_{\cN}^{-}$ is not manifest in the factorized form (\ref{eq:ANfch})
and due to the fact that $\hE(x)\neq E(x)$ the factorized form is
in appearance not invariant:
\begin{align}
P_{\cN}^{-}[\hW,\hE]&=\prod_{k=0}^{\cN-1}\left(\frac{\rmd}{\rmd x}
 +\hW(x)+\frac{\cN-1-2k}{2}\hE(x)\right)\notag\\
&=\prod_{k=0}^{\cN-1}\left(\frac{\rmd}{\rmd x}+W(x)+\frac{\cN-1-2k}{2}
 \left(E(x)-\frac{2\gamma z'(x)}{\gamma z(x)-\alpha}\right)\right).
\label{eq:traPN}
\end{align}
The fact that $P_{\cN}^{-}$ is also invariant under the $GL(2,\bbC)$
transformations
\begin{align}
P_{\cN}^{-}[\hW,\hE]=P_{\cN}^{-}[W,E],
\end{align}
proved in Ref.~\cite{Ta03a}, despite the non-invariance in appearance
for $\gamma\neq0$, indicates that the type A $\cN$-fold supercharge
admits a one-parameter family of different factorizations characterized
by the parameter $\alpha/\gamma$.

\section{Intermediate Hamiltonians for $\cN=2$}
\label{sec:inHam}

{}From now on, we shall restrict ourselves to the case of $\cN=2$.
The type A $\cN$-fold SUSY systems (\ref{eq:AHams}) and (\ref{eq:ANfch})
for $\cN=2$ read
\begin{align}
2H^{\pm}&=-\frac{\rmd^{2}}{\rmd x^{2}}+W(x)^{2}
 -\frac{E'(x)}{2}+\frac{E(x)^{2}}{4}-2R\pm 2W'(x),
\label{eq:A2pm}\\
P_{2}^{-}=P_{21}^{-}P_{22}^{-}&=\frac{\rmd^{2}}{\rmd x^{2}}+2W(x)
 \frac{\rmd}{\rmd x}+W'(x)+W(x)^{2}+\frac{E'(x)}{2}-\frac{E(x)^{2}}{4},
\label{eq:A2fch}
\end{align}
where
\begin{align}
P_{21}^{-}=\frac{\rmd}{\rmd x}+W(x)-\frac{E(x)}{2},\qquad
P_{22}^{-}=\frac{\rmd}{\rmd x}+W(x)+\frac{E(x)}{2}.
\label{eq:defQ-}
\end{align}
The superHamiltonian $\bH_{\!2}$ and the type A 2-fold supercharges
$\bQ_{2}^{\pm}$ introduced with the ordinary fermionic variables
$\psi^{\pm}$ as
\begin{align}
\bH_{\!2}=H^{-}\psi^{-}\psi^{+}+H^{+}\psi^{+}\psi^{-},\qquad
 \bQ_{2}^{\pm}=P_{2}^{\mp}\psi^{\pm},
\end{align}
satisfy the type A 2-fold superalgebra~\cite{Ta03a}:
\begin{align}
\bigl[\bQ_{2}^{\pm},\bH_{\!2}\bigr]=\bigl\{\bQ_{2}^{\pm},
 \bQ_{2}^{\pm}\bigr\}=0,\qquad\bigl\{\bQ_{2}^{-},\bQ_{2}^{+}
 \bigr\}=4(\bH_{\!2}+R)^{2}+4b_{0}b_{2}-b_{1}^{\,2}.
\label{eq:A2alg}
\end{align}
In the expanded form of the type A 2-fold supercharge components
(\ref{eq:A2fch}), its invariance under the $GL(2,\bbC)$ transformations
is now manifest by applying (\ref{eq:traW}) and (\ref{eq:traE2}):
\begin{align}
P_{2}^{-}[\hW,\hE]=P_{2}^{-}[W,E].
\label{eq:invP2}
\end{align}
However, each factor of the type A $\cN$-fold supercharge in the factorized
form is not invariant since $\widehat{E}(x)\neq E(x)$ as shown in
(\ref{eq:traPN}), and thus we generally have
\begin{align}
\hP_{21}^{-}\equiv P_{21}^{-}[\hW,\hE]\neq P_{21}^{-}[W,E],\qquad
 \hP_{22}^{-}\equiv P_{22}^{-}[\hW,\hE]\neq P_{22}^{-}[W,E].
\label{eq:ninv}
\end{align}
Next, we introduce another Hamiltonian $H^{\rmi1}$, which we shall call
an \emph{intermediate} Hamiltonian, as
\begin{align}
P_{22}^{-}H^{-}=H^{\rmi1}P_{22}^{-},\qquad
 P_{21}^{-}H^{\rmi1}=H^{+}P_{21}^{-},
\label{eq:defH0}
\end{align}
which are compatible with (\ref{eq:inter}). It is evident that
$H^{\rmi1}$ is in general not invariant under the $GL(2,\bbC)$
transformation in contrast with $H^{\pm}$ due to the fact that
both of $P_{21}^{-}$ and $P_{22}^{-}$ which intertwine $H^{\rmi1}$
with $H^{\pm}$ have no invariance (\ref{eq:ninv}). Hence, we can
expect a family of intermediate Hamiltonians for each given type A
2-fold SUSY system.
Needless to say, however, we do not always have such an intermediate
Hamiltonian for a given system. The necessary and sufficient conditions
for its existence are that there exist two constants $C_{22}$ and
$C_{21}$ such that $H^{\pm}$ and $H^{\rmi1}$ are expressed as
(see, e.g., Refs.~\cite{CKS95,Ju96,Ba00})
\begin{align}
\begin{split}
2H^{-}&=P_{22}^{+}P_{22}^{-}+2C_{22},\quad 2H^{+}=P_{21}^{-}
 P_{21}^{+}+2C_{21},\\
2H^{\rmi1}&=P_{22}^{-}P_{22}^{+}+2C_{22}=P_{21}^{+}P_{21}^{-}+2C_{21},
\label{eq:cond0}
\end{split}
\end{align}
where $P_{ij}^{+}$ are the transpositions of $P_{ij}^{-}$, that
is,\footnote{Note that we do not assume the reality of the functions
$W(x)$ and $E(x)$.}
\begin{align}
P_{21}^{+}=-\frac{\rmd}{\rmd x}+W(x)-\frac{E(x)}{2},\qquad
P_{22}^{+}=-\frac{\rmd}{\rmd x}+W(x)+\frac{E(x)}{2}.
\label{eq:defQ+}
\end{align}
More explicitly, the conditions (\ref{eq:cond0}) read
\begin{subequations}
\label{eqs:cond1}
\begin{align}
2H^{-}&=-\frac{\rmd^{2}}{\rmd x^{2}}+W(x)^{2}+E(x)W(x)
 +\frac{E(x)^{2}}{4}-\frac{E'(x)}{2}-W'(x)+2C_{22},\\
2H^{\rmi1}&=-\frac{\rmd^{2}}{\rmd x^{2}}+W(x)^{2}+E(x)W(x)
 +\frac{E(x)^{2}}{4}+\frac{E'(x)}{2}+W'(x)+2C_{22}\notag\\
&=-\frac{\rmd^{2}}{\rmd x^{2}}+W(x)^{2}-E(x)W(x)+\frac{E(x)^{2}}{4}
 +\frac{E'(x)}{2}-W'(x)+2C_{21},
\label{eq:H0}\\
2H^{+}&=-\frac{\rmd^{2}}{\rmd x^{2}}+W(x)^{2}-E(x)W(x)
 +\frac{E(x)^{2}}{4}-\frac{E'(x)}{2}+W'(x)+2C_{21}.
\end{align}
\end{subequations}
{}From Eqs.~(\ref{eq:A2pm}) and (\ref{eqs:cond1}), the necessary and
sufficient conditions reduce to
\begin{align}
W'(x)+E(x)W(x)=-2R-2C_{22}=C_{21}-C_{22}=2C_{21}+2R.
\label{eq:cond2}
\end{align}
Noting the relation
\begin{align}
W'(x)+E(x)W(x)=-Q'(z),
\end{align}
which easily follows from (\ref{eq:defE}) and (\ref{eq:defAQ}), we
find that the latter conditions (\ref{eq:cond2}) are equivalent to
\begin{align}
Q(z)=(C_{22}-C_{21})z+b_{0},\qquad -2R=C_{22}+C_{21},
\end{align}
with $b_{0}$ being another constant. We recall that for the most general
type A $\cN$-fold SUSY systems for all $\cN\geq2$, $Q(z)$ is given by
a polynomial of at most second-degree (\ref{eq:Qofz}).
Hence, a given type A 2-fold SUSY system (\ref{eq:A2pm}) admits an
intermediate Hamiltonian $H^{\rmi1}$ satisfying Eq.~(\ref{eq:defH0}) if
and only if
\begin{align}
b_{2}=0,\quad b_{1}=C_{22}-C_{21},\quad -2R=C_{22}+C_{21}.
\label{eq:cond3}
\end{align}
The last two conditions in (\ref{eq:cond3}) just determine these
constants for the given values of $b_{1}$ and $R$. Hence, only
the first condition in (\ref{eq:cond3}) is essential for the existence
of an intermediate Hamiltonian.

As was discussed previously, the type A 2-fold supercharge $P_{2}^{-}$
is invariant under the $GL(2,\bbC)$ transformation (\ref{eq:invP2})
while its factors $P_{22}^{-}$ and $P_{21}^{-}$ are not (\ref{eq:ninv})
for $\gamma\neq0$. As a consequence, the necessary and sufficient
conditions (\ref{eq:cond3}) for the existence of another intermediate
Hamiltonian $H^{\rmi2}$ after a $GL(2,\bbC)$ transformation are
accordingly changed as
\begin{align}
\hat{b}_{2}=0,\quad\hat{b}_{1}=\hat{C}_{22}-\hat{C}_{21},\quad
 -2R=\hat{C}_{22}+\hat{C}_{21},
\label{eq:cond3'}
\end{align}
where $\hat{C}_{22}$ and $\hat{C}_{21}$ are another set of constants.
Again, only the first condition in (\ref{eq:cond3'}) is essential for
the existence of another intermediate Hamiltonian $H^{\rmi2}$ after
the transformation. Under the fulfillment of the conditions
(\ref{eq:cond3'}), the original type A 2-fold SUSY Hamiltonians
$H^{\pm}$ and the new intermediate Hamiltonian $H^{\rmi2}$ are
expressed in terms of the transformed supercharges as
\begin{align}
\begin{split}
2H^{-}&=\hP_{22}^{+}\hP_{22}^{-}+2\hat{C}_{22},\quad
 2H^{+}=\hP_{21}^{-}\hP_{21}^{+}+2\hat{C}_{21},\\
2H^{\rmi2}&=\hP_{22}^{-}\hP_{22}^{+}+2\hat{C}_{22}=\hP_{21}^{+}
 \hP_{21}^{-}+2\hat{C}_{21},
\label{eq:Hi2}
\end{split}
\end{align}
where $\hP_{ij}^{+}$ are the transpositions of $\hP_{ij}^{-}$ which
were defined in (\ref{eq:ninv}), that is,
\begin{align}
\hP_{21}^{\pm}=\mp\frac{\rmd}{\rmd x}+\hW(x)-\frac{\hE(x)}{2},\qquad
\hP_{22}^{\pm}=\mp\frac{\rmd}{\rmd x}+\hW(x)+\frac{\hE(x)}{2}.
\label{eq:hQpm}
\end{align}
It is now clear that a type A 2-fold SUSY system which satisfies the
conditions (\ref{eq:cond3}) and thus admits an intermediate Hamiltonian
$H^{\rmi1}$ also admits another different intermediate Hamiltonian
$H^{\rmi2}$ after a $GL(2,\bbC)$ transformation with $\gamma\neq0$ if
and only if the conditions (\ref{eq:cond3'}) are simultaneously fulfilled
in addition to (\ref{eq:cond3}). It essentially means the satisfaction
of $b_{2}=\hat{b}_{2}=0$. From the transformation formula
(\ref{eq:trabi}) we immediately see that it is only possible for the
$GL(2,\bbC)$ transformation with $\gamma\neq0$ which satisfies
\begin{align}
\alpha b_{1}+\gamma b_{0}=0.
\label{eq:cond4}
\end{align}
Let us first consider the case when $b_{1}=0$. In this case we can assume
that $b_{0}\neq0$; otherwise $Q(z)=0$ since $b_{2}=0$ is already assumed
in order to meet the condition (\ref{eq:cond3}), and from (\ref{eq:VAQ})
$V^{-}(x)=V^{+}(x)$ which means that the system is trivial as 2-fold
SUSY. But for $b_{1}=0$ and $b_{0}\neq0$ there is no solution to the
equation (\ref{eq:cond4}) except for $\gamma=0$. But for any $GL(2,\bbC)$
transformation with $\gamma=0$, the function $E(x)$ is invariant by
Eq.~(\ref{eq:traE}) and so is the intermediate Hamiltonian $H^{\rmi1}$.
Hence in the case of $b_{1}=0$ the factorization of type A 2-fold
supercharge admitting intermediate Hamiltonians is unique. On the other
hand, in the case of $b_{1}\neq0$ the solution to the equation
(\ref{eq:cond4}) is given by
\begin{align}
\alpha/\gamma=-b_{0}/b_{1}.
\label{eq:cond5}
\end{align}
As was shown previously, any type A 2-fold supercharge admits
one-parameter family of factorizations $P_{2}^{-}=P_{21}^{-}P_{22}^{-}$
characterized by the parameter $\alpha/\gamma$. In addition, any type A
2-fold SUSY system with $b_{2}=0$ has at least one intermediate
Hamiltonian $H^{\rmi1}$.
Then, the result (\ref{eq:cond5}) tells us that if the system further
satisfies $b_{1}\neq0$, it can admit a one and only one additional and
different intermediate Hamiltonian $H^{\rmi2}$ at the one point
(\ref{eq:cond5}) in the parameter space of $\alpha/\gamma\in\bbC$.
In Table~\ref{tb:numb}, we summarize the results.
\begin{table}
\begin{center}
\tabcolsep=10pt
\begin{tabular}{cc}
\hline
Conditions & Number of $H^{\rmi}$\\
\hline
$b_{2}\neq0$ & 0\\
$b_{2}=b_{1}=0$ & 1\\
$b_{2}=0$, $b_{1}\neq0$ & 2\\
\hline
\end{tabular}
\caption{The admissible numbers of different intermediate Hamiltonians
in type A 2-fold supersymmetry.}
\label{tb:numb}
\end{center}
\end{table}

\section{Second-Order Parasupersymmetry}
\label{sec:psusy}

In the previous section, we have just verified that a type A 2-fold SUSY
system $(H^{\pm}, P_{2}^{-}=P_{21}^{-}P_{22}^{-})$ admits (at least) one
intermediate Hamiltonian $H^{\rmi1}$ if and only if the condition
$b_{2}=0$ holds. In this section, we shall further show that any such
a system can possess an additional symmetry, namely, parasupersymmetry
of order 2 introduced in Ref.~\cite{RS88}.
Indeed, for a given such type A 2-fold SUSY system we can define
a triple of operators $(\bH_{\!\rmP},\bQ_{\rmP}^{\pm})$ by
\begin{subequations}
\label{eqs:pss2}
\begin{align}
\bH_{\!\rmP}&=H^{-}(\psi_{\rmP}^{-})^{2}(\psi_{\rmP}^{+})^{2}+H^{\rmi1}
 (\psi_{\rmP}^{+}\psi_{\rmP}^{-}-(\psi_{\rmP}^{+})^{2}(\psi_{\rmP}^{-}
 )^{2})+H^{+}(\psi_{\rmP}^{+})^{2}(\psi_{\rmP}^{-})^{2},\\
\bQ_{\rmP}^{-}&=P_{22}^{+}(\psi_{\rmP}^{-})^{2}\psi_{\rmP}^{+}
 +P_{21}^{+}\psi_{\rmP}^{+}(\psi_{\rmP}^{-})^{2},\quad\bQ_{\rmP}^{+}
 =P_{22}^{-}\psi_{\rmP}^{-}(\psi_{\rmP}^{+})^{2}+P_{21}^{-}
 (\psi_{\rmP}^{+})^{2}\psi_{\rmP}^{-},
\end{align}
\end{subequations}
where $\psi_{\rmP}^{\pm}$ are parafermions of order 2
satisfying~\cite{Ta07a}
\begin{align}
(\psi_{\rmP}^{\pm})^{2}\neq0,\quad(\psi_{\rmP}^{\pm})^{3}=0,\quad
 \bigl\{\psi_{\rmP}^{-},\psi_{\rmP}^{+}\bigr\}+\bigl\{(\psi_{\rmP}^{-}
 )^{2},(\psi_{\rmP}^{+})^{2}\bigr\}=2I.
\end{align}
Then, using Eq.~(\ref{eq:cond0}) and the parafermionic algebra of order
2 in Ref.~\cite{Ta07a} we can show that the triple $(\bH_{\!\rmP},
\bQ_{\rmP}^{\pm})$ defined as (\ref{eqs:pss2}) satisfies the second-order
paraSUSY relations in Ref.~\cite{RS88}:
\begin{align}
(\bQ_{\rmP}^{\pm})^{2}\neq0,\quad(\bQ_{\rmP}^{\pm})^{3}=0,\quad
 \bigl[\bQ_{\rmP}^{\pm},\bH_{\!\rmP}\bigr]=0,
\label{eq:para1}\\
(\bQ_{\rmP}^{\pm})^{2}\bQ_{\rmP}^{\mp}+\bQ_{\rmP}^{\pm}\bQ_{\rmP}^{\mp}
 \bQ_{\rmP}^{\pm}+\bQ_{\rmP}^{\mp}(\bQ_{\rmP}^{\pm})^{2}
 =4\bQ_{\rmP}^{\pm}\bH_{\!\rmP},
\label{eq:para2}
\end{align}
if and only if the constants $C_{ij}$ in (\ref{eq:cond0}) satisfy
\begin{align}
C_{22}=-C_{21}=b_{1}/2,
\label{eq:cpara}
\end{align}
and thus in particular $R=0$ by (\ref{eq:cond3}). Hence, we conclude
that any type A 2-fold SUSY quantum system with (at least) one
intermediate Hamiltonian also has paraSUSY of order 2 when $R=0$.
The additional restriction $R=0$ arises since one of the paraSUSY
conditions (\ref{eq:para2}) is not invariant under any constant shift
of $\bH_{\!\rmP}$.
Furthermore, as was shown in Ref.~\cite{Ta07a} this type of realization
of second-order paraSUSY systems admits an additional novel nonlinear
relation as the following (cf., Eq.~(6.66) in the latter reference):
\begin{align}
(\bQ_{\rmP}^{-})^{2}(\bQ_{\rmP}^{+})^{2}+\bQ_{\rmP}^{\pm}
 (\bQ_{\rmP}^{\mp})^{2}\bQ_{\rmP}^{\pm}+(\bQ_{\rmP}^{+})^{2}
 (\bQ_{\rmP}^{-})^{2}=4(\bH_{\!\rmP})^{2}-b_{1}^{\,2},
\label{eq:nlrel}
\end{align}
which can be regarded as a generalized (type A) 2-fold superalgebra.
In fact, on the one hand we immediately have from the paraSUSY relations
in (\ref{eq:para1})
\begin{align}
\bigl\{(\bQ_{\rmP}^{\pm})^{2},(\bQ_{\rmP}^{\pm})^{2}\bigr\}=0,\qquad
 \bigl[(\bQ_{\rmP}^{\pm})^{2},\bH_{\!\rmP}\bigr]=0,
\label{eq:salg1}
\end{align}
while on the other hand the nonlinear relation (\ref{eq:nlrel})
reduces, in the subsector with the parafermion number zero and two, to
\begin{align}
\bigl\{(\bQ_{\rmP}^{-})^{2},(\bQ_{\rmP}^{+})^{2}\bigr\}
 =4(\bH_{\!\rmP})^{2}-b_{1}^{\,2}.
\label{eq:salg2}
\end{align}
Then, the commutation and anti-commutation relations (\ref{eq:salg1})
and (\ref{eq:salg2}) are, under the assumed condition $b_{2}=0$
and $R=0$, entirely identical with the type A 2-fold superalgebra
(\ref{eq:A2alg}) with the trivial identification of the type A 2-fold
supercharges $\bQ_{2}^{\pm}$ with $(\bQ_{\rmP}^{\pm})^{2}$ and with
the observation that in the subsector $\bH_{\!\rmP}$ is essentially
identical with $\bH_{\!2}$. The relation between type A 2-fold SUSY
and second-order paraSUSY was briefly referred to in Ref.~\cite{Ta07a}.
Here we have firstly shown the necessary and sufficient conditions
for a type A 2-fold SUSY system to admit simultaneously second-order
paraSUSY, namely, Eqs.~(\ref{eq:cond3}) and (\ref{eq:cpara}). Finally,
it is evident that we can construct two sets of second-order paraSUSY
systems whenever a type A 2-fold SUSY system has two different
intermediate Hamiltonians.

\section{An Application to the generalized P\"{o}schl--Teller potential}
\label{sec:appli}

As an application of the general framework discussed in the previous
two sections, we shall reconstruct the BQR SSUSY model (\ref{eq:BQRs})%
--(\ref{eq:BQRe}) and its generalization which preserves all the SUSY
and SSUSY structure therein. To this end, let us first choose
the change of variable $z=z(x)$ which determines the relation
between physical Hamiltonians and gauged ones as
\begin{align}
z(x)&=-(\sinh x)^{-2B}\left(\tanh\frac{x}{2}\right)^{2A+1}
 \notag\\
&=-2^{-2B}\left(\sinh\frac{x}{2}\right)^{2A-2B+1}
 \left(\cosh\frac{x}{2}\right)^{-(2A+2B+1)},
\end{align}
where $A$ and $B$ are both constants. The function $A(z)=z'(x)^{2}/2$
defined in (\ref{eq:defAQ}) and its derivatives with respect to $z$
are in general transcendental functions of $z$. Explicitly, they read
\begin{align}
z'(x)&=-z(x)\frac{2B\cosh x-2A-1}{\sinh x},
\label{eq:z'x}\\
A(z)&=\frac{z(x)^{2}(2B\cosh x-2A-1)^{2}}{2\sinh^{2}x},
\label{eq:Aofz}\\
A'(z)&=z(x)\left(4B^{2}-\frac{\alpha_{1}^{+}\cosh x-\alpha_{2}^{+}}{
 \sinh^{2}x}\right),\\
A''(z)&=4B^{2}-\frac{\beta_{1}\cosh x-\beta_{2}}{\sinh^{2}x}
 -\frac{2A+1}{2B\cosh x-2A-1},
\end{align}
where $\alpha_{i}^{+}$ and $\beta_{i}$ are all constants given by
\begin{alignat}{2}
\alpha_{1}^{\pm}&=(2A+1)(4B\pm1),&\qquad
 \alpha_{2}^{\pm}&=(2A+1)^{2}+2B(2B\pm1),
\label{eq:alpha}\\
\beta_{1}&=(2A+1)(4B+3),&\beta_{2}&=(2A+1)^{2}+2(B+1)(2B+1).
\end{alignat}
Substituting them into the most general form of a pair of type A 2-fold
SUSY potentials, Eq.~(\ref{eq:VAQ}) with $\cN=2$, we obtain
\begin{align}
V^{\pm}(x)=&\,\frac{Q(z(x))^{2}\sinh^{2}x}{2z(x)^{2}(2B\cosh x-2A-1)^{2}}
+\frac{4(2A+1)B\cosh x+(2A+1)^{2}-12B^{2}}{8(2B\cosh x-2A-1)^{2}}\notag\\
&-\frac{(2A+1)B\cosh x-A(A+1)-B^{2}}{2\sinh^{2}x}+\frac{B^{2}}{2}
 -R\notag\\
&\pm\frac{Q(z(x))(4B^{2}\sinh^{2}x-\alpha_{1}^{+}\cosh x+\alpha_{2}^{+}
 )}{z(x)(2B\cosh x-2A-1)^{2}}\mp Q'(z(x)),
\label{eq:Vpm}
\end{align}
where $Q(z)$ is a polynomial of at most second-degree given as in
(\ref{eq:Qofz}). The function $W(x)$ characterizing the type A system
in this case reads
\begin{align}
W(x)=-\frac{Q(z(x))}{z'(x)}=\frac{Q(z(x))\sinh x}{z(x)(2B\cosh x-2A-1)}.
\label{eq:Wx2}
\end{align}
The other function $E(x)$ defined through the relation (\ref{eq:defE}),
which also characterizes the type A system, is calculated as
\begin{align}
E(x)=\frac{z''(x)}{z'(x)}=-\frac{(2B+1)\cosh x-2A-1}{\sinh x}
 +\frac{2B\sinh x}{2B\cosh x-2A-1}.
\label{eq:Ex}
\end{align}
Next, let us consider the case when the type A 2-fold SUSY system admits
an intermediate Hamiltonian, namely, $b_{2}=0$. From (\ref{eq:Wx2}) and
(\ref{eq:Ex}) we have
\begin{align}
W'(x)=&\,\frac{Q(z(x))}{z(x)}-Q'(z(x))
 -\frac{(2A+1)\cosh x-2B}{z(x)(2B\cosh x-2A-1)^{2}}Q(z(x)),\\
E'(x)=&-\frac{2(2A+1)B\cosh x-4B^{2}}{(2B\cosh x-2A-1)^{2}}
 -\frac{(2A+1)\cosh x-2B-1}{\sinh^{2}x},\\
E(x)^{2}=&\,4B^{2}+\frac{(2A+1)^{2}-4B^{2}}{(2B\cosh x-2A-1)^{2}}\notag\\
 &-\frac{2(2A+1)(2B+1)\cosh x-(2A+1)^{2}-(2B+1)^{2}}{\sinh^{2}x}.
\label{eq:Ex^2}
\end{align}
Substituting (\ref{eq:Wx2})--(\ref{eq:Ex^2}) into (\ref{eq:H0}), using
the relations (\ref{eq:cond3}) among the constants, and noting that
$Q'(z)=b_{1}$ when $b_{2}=0$, we obtain the intermediate potential
$V^{\rmi1}(x)$ as
\begin{align}
V^{\rmi1}(x)=&\,\frac{Q(z(x))^{2}\sinh^{2}x}{2z(x)^{2}(2B\cosh x
 -2A-1)^{2}}-\frac{4(2A+1)B\cosh x-(2A+1)^{2}-4B^{2}}{8(2B\cosh x
 -2A-1)^{2}}\notag\\
&-\frac{(2A+1)(B+1)\cosh x-A(A+1)-(B+1)^{2}}{2\sinh^{2}x}
 +\frac{B^{2}}{2}-R.
\label{eq:Vi1}
\end{align}
Next, we shall consider the case when the system admits another different
intermediate Hamiltonian $H^{\rmi2}$, namely, $b_{1}\neq0$. The
$GL(2,\bbC)$ transformation which takes the type A 2-fold supercharge to
another factorization for which $H^{\rmi2}$ exists must satisfy
the condition (\ref{eq:cond5}). The parameter $\delta$ does not play an
important role in our context, so we set $\delta=0$ without any loss of
generality. But in this case $\beta$ cannot be $0$ otherwise $\Delta=0$.
Thus, we fix the parameters as
\begin{align}
\alpha/\gamma=-b_{0}/b_{1}\equiv -z_{0},\qquad\beta/\gamma=-1,
 \qquad\delta=0.
\label{eq:parav}
\end{align}
In other words, we choose the following $GL(2,\bbC)$ transformation on
the variable $z(x)$:
\begin{align}
w(x)=-\frac{1}{z(x)+z_{0}}=\frac{1}{(\sinh x)^{-2B}(\tanh x/2
 )^{2A+1}-z_{0}}.
\end{align}
The function $W(x)$ is invariant under the transformation (see,
Eq.~(\ref{eq:traW})) while $E(x)$ is transformed according to
(\ref{eq:traE}) as
\begin{align}
\hE(x)=&\,E(x)-\frac{2z'(x)}{z(x)+z_{0}}=E(x)+\frac{2z(x)}{z(x)+z_{0}}
 \frac{2B\cosh x-2A-1}{\sinh x}\notag\\
=&\,\frac{[(2B-1)z(x)-(2B+1)z_{0}]\cosh x-(2A+1)(z(x)-z_{0})}{
 (z(x)+z_{0})\sinh x}\notag\\
&+\frac{2B\sinh x}{2B\cosh x-2A-1}.
\label{eq:hatE}
\end{align}
Noting the relation $Q(z)=b_{1}(z+z_{0})$ when $b_{2}=0$, we have
\begin{align}
\hE(x)\hW(x)=E(x)W(x)+2b_{1}.
\end{align}
{}From the second expression of $\hE(x)$ in (\ref{eq:hatE}), we obtain
the following formulas:
\begin{align}
\hE'(x)=&\,E'(x)+\frac{2z(x)}{z(x)+z_{0}}
 \frac{(2A+1)\cosh x-2B}{\sinh^{2}x}\notag\\
&-\frac{2z_{0}z(x)}{(z(x)+z_{0})^{2}}
 \frac{(2B\cosh x-2A-1)^{2}}{\sinh^{2}x},\\
\hE(x)^{2}=&\,E(x)^{2}+\frac{8Bz(x)}{z(x)+z_{0}}-\frac{4z(x)\cosh x
 (2B\cosh x-2A-1)}{(z(x)+z_{0})\sinh^{2}x}\notag\\
&-\frac{4z_{0}z(x)(2B\cosh x-2A-1)^{2}}{(z(x)+z_{0})^{2}\sinh^{2}x}.
\label{eq:hE^2}
\end{align}
Substituting (\ref{eq:hatE})--(\ref{eq:hE^2}) into Eq.~(\ref{eq:Hi2})
and noting that $\hat{b}_{1}=-b_{1}$ by the transformation formula
(\ref{eq:trabi}) in our choice of the parameters (\ref{eq:parav}), we
obtain for $H^{\pm}$ the same potentials as the ones in (\ref{eq:Vpm}),
as they should be, while for the other intermediate Hamiltonian
$H^{\rmi2}$ the following form of the potential:
\begin{align}
V^{\rmi2}(x)=&\,V^{\rmi1}(x)+\frac{z(x)^{2}[(2A+1)\cosh x-2B]}{
 (z(x)+z_{0})^{2}\sinh^{2}x}\notag\\
&+\frac{z_{0}z(x)(\alpha_{1}^{+}\cosh x-\alpha_{2}^{+})}{(z(x)+z_{0})^{2}
 \sinh^{2}x}-\frac{4B^{2}z_{0}z(x)}{(z+z_{0})^{2}}.
\end{align}
If we substitute (\ref{eq:Vi1}) for $V^{\rmi1}(x)$ into the above, we
finally obtain the full expression of $V^{\rmi2}(x)$ as
\begin{align}
V^{\rmi2}(x)=&\,\frac{Q(z(x))^{2}\sinh^{2}x}{2z(x)^{2}(2B\cosh x-
 2A-1)^{2}}-\frac{4(2A+1)B\cosh x-(2A+1)^{2}-4B^{2}}{8(2B\cosh x
 -2A-1)^{2}}\notag\\
&-\frac{(2A+1)(B-1)\cosh x-A(A+1)-(B-1)^{2}}{2\sinh^{2}x}
 +\frac{B^{2}}{2}-R\notag\\
&+\frac{z_{0}z(x)(\alpha_{1}^{-}\cosh x-\alpha_{2}^{-})}{
 (z(x)+z_{0})^{2}\sinh^{2}x}
-\frac{z_{0}^{\,2}[(2A+1)\cosh x-2B]}{(z(x)+z_{0})^{2}}
 -\frac{4B^{2}z_{0}z(x)}{(z(x)+z_{0})^{2}},
\label{eq:Vi2}
\end{align}
where $\alpha_{i}^{-}$ are defined in (\ref{eq:alpha}). We are now
in a position to show that the type A 2-fold system with the two
intermediate Hamiltonians (\ref{eq:Vpm}), (\ref{eq:Vi1}), and
(\ref{eq:Vi2}) contains as a special case the BQR SSUSY model
(\ref{eq:BQRs})--(\ref{eq:BQRe}). For the purpose, let put $b_{0}=0$
and $b_{1}=bB$. In this case, $Q(z)=bBz$ and $z_{0}=b_{0}/b_{1}=0$.
Then, the 2-fold SUSY pair of the potentials (\ref{eq:Vpm}) and the
two intermediate potentials (\ref{eq:Vi1}) and (\ref{eq:Vi2})
reduce to, respectively,
\begin{align}
V^{\pm}(x)=&\,\frac{4(b^{2}+1)(2A+1)B\cosh x-(b^{2}-1)(2A+1)^{2}
 -4(b^{2}+3)B^{2}}{8(2B\cosh x-2A-1)^{2}}\notag\\
&-\frac{(2A+1)B\cosh x-A(A+1)-B^{2}}{2\sinh^{2}x}+\frac{b^{2}}{8}
 +\frac{B^{2}}{2}-R\notag\\
&\mp bB\frac{(2A+1)\cosh x-2B}{(2B\cosh x-2A-1)^{2}},
\label{eq:Vpm'}
\end{align}
and
\begin{align}
V^{\rmi1}(x)=&\,(b^{2}-1)\frac{4(2A+1)B\cosh x-(2A+1)^{2}-4B^{2}}{
 8(2B\cosh x-2A-1)^{2}}\notag\\
&-\frac{(2A+1)(B+1)\cosh x-A(A+1)-(B+1)^{2}}{\sinh^{2}x}
 +\frac{b^{2}}{8}+\frac{B^{2}}{2}-R,
\label{eq:Vi1'}
\end{align}
and
\begin{align}
V^{\rmi2}(x)=&\,(b^{2}-1)\frac{4(2A+1)B\cosh x-(2A+1)^{2}-4B^{2}}{
 8(2B\cosh x-2A-1)^{2}}\notag\\
&-\frac{(2A+1)(B-1)\cosh x-A(A+1)-(B-1)^{2}}{\sinh^{2}x}
 +\frac{b^{2}}{8}+\frac{B^{2}}{2}-R.
\label{eq:Vi2'}
\end{align}
The components of supercharges $P_{ij}^{\pm}$ and $\hP_{ij}^{\pm}$
given by (\ref{eq:defQ-}), (\ref{eq:defQ+}), and (\ref{eq:hQpm}) in
this case read
\begin{align}
P_{21}^{\pm}&=\mp\frac{\rmd}{\rmd x}+\frac{(2B+1)\cosh x-2A-1}{2\sinh x}
 +\frac{(b-1)B\sinh x}{2B\cosh x-2A-1},\\
P_{22}^{\pm}&=\mp\frac{\rmd}{\rmd x}-\frac{(2B+1)\cosh x-2A-1}{2\sinh x}
 +\frac{(b+1)B\sinh x}{2B\cosh x-2A-1},\\
\hP_{21}^{\pm}&=\mp\frac{\rmd}{\rmd x}-\frac{(2B-1)\cosh x-2A-1}{2\sinh x}
 +\frac{(b-1)B\sinh x}{2B\cosh x-2A-1},\\
\hP_{22}^{\pm}&=\mp\frac{\rmd}{\rmd x}+\frac{(2B-1)\cosh x-2A-1}{2\sinh x}
 +\frac{(b+1)B\sinh x}{2B\cosh x-2A-1}.
\end{align}
It is now easy to see that the BQR SSUSY model is realized when $b=-1$.
Indeed, we have the following correspondences when $b=-1$:
\begin{align}
2V^{-}(x)=V_{A,B}(x)+\frac{1}{4}+B^{2}-2R,\quad
 2V^{+}(x)=V_{A,B,\textrm{ext}}(x)+\frac{1}{4}+B^{2}-2R,\\
2V^{\rmi1}(x)=V_{A,B+1}(x)+\frac{1}{4}+B^{2}-2R,\quad
 2V^{\rmi2}(x)=V_{A,B-1}(x)+\frac{1}{4}+B^{2}-2R,\\
P_{21}^{-}\ \textrm{or}\ \hP_{21}^{-}=\hat{A},\quad
 P_{21}^{+}\ \textrm{or}\ \hP_{21}^{+}=\hat{A}^{\dagger},\quad
 P_{22}^{-}\ \textrm{or}\ \hP_{22}^{-}=\hat{B},\quad
 P_{22}^{+}\ \textrm{or}\ \hP_{22}^{+}=\hat{B}^{\dagger},
\end{align}
and in particular $P_{2}^{-}=P_{21}^{-}P_{22}^{-}=\hP_{21}^{-}\hP_{22}^{-}
=\hat{A}\hat{B}$. The relations among the constants are given by
\begin{align}
\bar{c}=4C_{22}=-4C_{21}=-2B\quad\textrm{or}
 \quad\bar{c}=4\hat{C}_{22}=-4\hat{C}_{21}=2B,\\
-\tilde{E}+\frac{\bar{c}}{2}=-E-\frac{\bar{c}}{2}
 =B^{2}+\frac{1}{4},\qquad R=0.
\end{align}
The last equality $R=0$ means from the results in Section~\ref{sec:psusy}
that the BQR SSUSY model also has second-order paraSUSY.
We note that the reason why the BQR SSUSY model is realized as the
particular case $b_{2}=b_{0}=0$ of the most general type A 2-fold SUSY
is the same as the one discussed in Ref.~\cite{GT06}, Section~5.

\section{Discussion and Summary}
\label{sec:discus}

In this article, we have investigated in detail under what conditions
type A $\cN$-fold SUSY systems can have intermediate Hamiltonians in
the case of $\cN=2$. It turns out that although type A 2-fold
supercharge admits a one-parameter family of factorization into
product of two first-order linear differential operators due to
the underlying $GL(2,\bbC)$ symmetry, at most two different
intermediate Hamiltonians are admissible. As a by product of the
studies, we have also obtained the necessary and sufficient conditions
for a type A 2-fold SUSY system to possess paraSUSY of order 2 as well.
When it is the case, the type A 2-fold superalgebra together with
the second-order parasuperalgebra constitute a generalized 2-fold
superalgebra. As a demonstration of the general arguments, we have
constructed the generalized P\"{o}schl--Teller potentials which
are components of type A 2-fold SUSY with two intermediate Hamiltonians
and reduce to the BQR SSUSY model in a particular case.

As for the concept like the reducibility in Ref.~\cite{AICD95},
the present investigations indicate that it would be more natural
and useful to classify higher-order intertwining operators according
to the existence and the number of intermediate Hamiltonians as has
been done in Table~\ref{tb:numb}. After employing the latter
classification scheme, we can further classify them according to
the properties of the intermediate Hamiltonians such as Hermiticity,
$\cP\cT$ symmetry, and so on.

Regarding the generalized P\"{o}schl--Teller potentials constructed in
Section~\ref{sec:appli}, it is worth noticing that the framework of
$\cN$-fold SUSY works well even when the function $A(z)$, which
controls the change of variable $z=z(x)$ from the physical coordinate
$x$ to the variable $z$ in the gauged space, is a transcendental
function of $z$ without destroying quasi-solvability. For all
$\cN\geq3$ cases type A $\cN$-fold SUSY requires the additional
condition (\ref{eq:condA}) so that $A(z)$ is allowed to be at most
a polynomial of fourth-degree in $z$, which results in the admissible
change of variable to be at most an elliptic function, see, e.g.,
Ref.~\cite{Ta03a}. For the $\cN=2$ case, on the other hand, there
are no such restrictions and, to the best of our knowledge, our
generalized P\"{o}schl--Teller potentials are the first quasi-solvable
examples where $A(z)$ is given by a transcendental function of $z$
as Eq.~(\ref{eq:Aofz}).

The analyses for $\cN=2$ carried out in this article are easily
generalized to the cases $\cN\geq3$, but we anticipate that richer
structure could emerge for the higher $\cN$ cases. In the case of
$\cN=3$, for instance, according to the factorization of type A 3-fold
supercharge $P_{3}^{-}=P_{31}^{-}P_{32}^{-}P_{33}^{-}$ we can consider
not only the case where intermediate Hamiltonians between $P_{31}^{-}$
and $P_{32}^{-}$ and between $P_{32}^{-}$ and $P_{33}^{-}$ both exist,
but also the cases where they exist only between the former place or
only between the latter place exclusively. It is also interesting to
study whether or not type A $\cN$-fold SUSY systems for higher $\cN$,
when they have intermediate Hamiltonians, can admit another symmetry.
The fact that in the case of $\cN=2$ they have second-order paraSUSY
indicates that they could have higher-order paraSUSY~\cite{To92,Kh92}
for $\cN\geq3$. Indeed, it was shown in Ref.~\cite{Ta07c} that a
certain realization of paraSUSY of order 3 also admits a generalized
3-fold superalgebra. Hence, at least in the case of $\cN=3$ we have
a reasonable basis to expect paraSUSY as an additional symmetry.
Other candidates might be quasi-paraSUSY introduced in
Ref.~\cite{Ta07a} and $\cN$-fold paraSUSY in Ref.~\cite{Ta07b}.

\begin{acknowledgments}
 This work (T.Tanaka) was partially supported by the National Cheng Kung
 University under the grant No.\ OUA:95-3-2-071.
\end{acknowledgments}



\bibliography{refsels}

\begin{thebibliography}{10}
\expandafter\ifx\csname url\endcsname\relax
  \def\url#1{{\tt #1}}\fi
\expandafter\ifx\csname urlprefix\endcsname\relax\def\urlprefix{URL }\fi
\providecommand{\eprint}[2][]{\url{#2}}

\bibitem{Wi81}
E.~Witten, Nucl. Phys. B 188 (1981) 513.

\bibitem{CKS95}
F.~Cooper, A.~Khare, and U.~Sukhatme, Phys. Rep. 251 (1995) 267.
\newblock \eprint{hep-th/9405029}.

\bibitem{Ju96}
G.~Junker, Supersymmetric {M}ethods in {Q}uantum and {S}tatistical {P}hysics
  (Springer, Berlin, 1996).

\bibitem{Ba00}
B.~K. Bagchi, Supersymmetry in {Q}uantum and {C}lassical {M}echanics (Chapman
  and Hall/CRC press, Florida, 2000).

\bibitem{Sc40a}
E.~Schr{\"{o}}dinger, Proc. R. Irish Acad. A 46 (1940) 9.

\bibitem{Da1882}
G.~Darboux, Comput. Rend. Acad. Sci. 94 (1882) 1456.

\bibitem{AIS93}
A.~A. Andrianov, M.~V. Ioffe, and V.~P. Spiridonov, Phys. Lett. A 174 (1993)
  273.
\newblock \eprint{hep-th/9303005}.

\bibitem{AST01b}
H.~Aoyama, M.~Sato, and T.~Tanaka, Nucl. Phys. B 619 (2001) 105.
\newblock \eprint{quant-ph/0106037}.

\bibitem{AS03}
A.~A. Andrianov and A.~V. Sokolov, Nucl. Phys. B 660 (2003) 25.
\newblock \eprint{hep-th/0301062}.

\bibitem{AICD95}
A.~A. Andrianov, M.~V. Ioffe, F.~Cannata, and J.~P. Dedonder, Int. J. Mod.
  Phys. A 10 (1995) 2683.
\newblock \eprint{hep-th/9404061}.

\bibitem{AIN95a}
A.~A. Andrianov, M.~V. Ioffe, and D.~N. Nishnianidze, Theor. Math. Phys. 104
  (1995) 1129.

\bibitem{BB98a}
C.~M. Bender and S.~Boettcher, Phys. Rev. Lett. 80 (1998) 5243.
\newblock \eprint{physics/9712001}.

\bibitem{CJT98}
F.~Cannata, G.~Junker, and J.~Trost, Phys. Lett. A 246 (1998) 219.
\newblock \eprint{quant-ph/9805085}.

\bibitem{BR00}
B.~Bagchi and R.~Roychoudhury, J. Phys. A: Math. Gen. 33 (2000) L1.
\newblock \eprint{quant-ph/9911104}.

\bibitem{ZCBR00}
M.~Znojil, F.~Cannata, B.~Bagchi, and R.~Roychoudhury, Phys. Lett. B 483 (2000)
  284.
\newblock \eprint{hep-th/0003277}.

\bibitem{Mi84}
B.~Mielnik, J. Math. Phys. 25 (1984) 3387.

\bibitem{Fe84}
D.~J. {Fern{\'a}ndez C.}, Lett. Math. Phys. 8 (1984) 337.

\bibitem{Zh87}
D.~Zhu, J. Phys. A: Math. Gen. 20 (1987) 4331.

\bibitem{Ku87}
C.~N. Kumar, J. Phys. A: Math. Gen. 20 (1987) 5397.

\bibitem{AF88}
N.~A. Alves and E.~Drigo Filho, J. Phys. A: Math. Gen. 21 (1988) 3215.

\bibitem{Fi88}
E.~Drigo Filho, J. Phys. A: Math. Gen. 21 (1988) L1025.

\bibitem{MRLB89}
A.~Mitra, P.~K. Roy, A.~Lahiri, and B.~Bagchi, Int. J. Theor. Phys. 28 (1989)
  911.

\bibitem{Qu08b}
C.~Quesne, J. Phys. A: Math. Theor. 41 (2008) 392001.
\newblock \eprint{arXiv:0807.4087 [quant-ph]}.

\bibitem{AST01a}
H.~Aoyama, M.~Sato, and T.~Tanaka, Phys. Lett. B 503 (2001) 423.
\newblock \eprint{quant-ph/0012065}.

\bibitem{Ta03a}
T.~Tanaka, Nucl. Phys. B 662 (2003) 413.
\newblock \eprint{hep-th/0212276}.

\bibitem{BQR08}
B.~Bagchi, C.~Quesne, and R.~Roychoudhury, Pramana J. Phys. 73 (2009) 337.
\newblock \eprint{arXiv:0812.1488 [quant-ph]}.

\bibitem{GT06}
A.~Gonz{\'a}lez-L{\'o}pez and T.~Tanaka, J. Phys. A: Math. Gen. 39 (2006) 3715.
\newblock \eprint{quant-ph/0602177}.

\bibitem{RS88}
V.~A. Rubakov and V.~P. Spiridonov, Mod. Phys. Lett. A 3 (1988) 1337.

\bibitem{Ta07a}
T.~Tanaka, Ann. Phys. 322 (2007) 2350.
\newblock \eprint{hep-th/0610311}.

\bibitem{To92}
M.~Tomiya, J. Phys. A: Math. Gen. 25 (1992) 4699.

\bibitem{Kh92}
A.~Khare, J. Phys. A: Math. Gen. 25 (1992) L749.

\bibitem{Ta07c}
T.~Tanaka, Ann. Phys. 322 (2007) 2682.
\newblock \eprint{hep-th/0612263}.

\bibitem{Ta07b}
T.~Tanaka, Mod. Phys. Lett. A 22 (2007) 2191.
\newblock \eprint{hep-th/0611008}.

\end{thebibliography}
\bibliographystyle{npb}



\end{document}